\documentclass[12pt]{article}

\usepackage{amsmath, amsthm, amssymb}

\usepackage{clrscode}
\usepackage[all]{xy}
\usepackage{lscape}       
\usepackage{booktabs}  

\usepackage{fullpage}
\usepackage{hyperref}
\usepackage{graphicx,url}
\usepackage[latin1]{inputenc}  
\usepackage{enumitem}  
\usepackage{subfigure}
\usepackage{setspace}  

\newtheorem{thm}{Theorem}[section] 
\newtheorem{prob}[thm]{Problem} %

\begin{document}
\onehalfspacing 

\thispagestyle{empty} 
\begin{center}
  {\LARGE \bf Finding steady-state solutions for ODE systems of zero, first and homogeneous second-order chemical reactions is NP-hard}\\

  \bigskip

  \href{mailto:marcelo.reis@butantan.gov.br}{Marcelo S. Reis}$^{1}$\\

  \bigskip

  $^1$ Center of Toxins, Immune-response and Cell Signaling (CeTICS);\\
Laborat\'orio Especial de Ciclo Celular, Instituto Butantan, S\~ao Paulo, Brazil\\

  \bigskip

  S\~ao Paulo, \today
\end{center}

\bigskip

\begin{abstract}
In the context of modeling of cell signaling pathways, a relevant step is finding steady-state solutions for ODE systems that describe the kinetics of a set of chemical reactions, especially sets composed of zero, first, and second-order reactions. To compute a steady-state solution, one must set the left-hand side of each ODE as zero, hence obtaining a system of non-negative, quadratic polynomial equations. If all second-order reactions are homogeneous in respect to their reactants, then the obtained quadratic polynomial equation system will also have univariate monomials. Although it is a well-known fact that finding a root of a quadratic polynomial equation system is a NP-hard problem, it is not so easy to find a readily available proof of NP-hardness for special cases like the aforementioned one. Therefore, we provide here a self-contained proof that finding a root of non-negative, with univariate monomials quadratic polynomial equation system (NUMQ-PES) is NP-hard. This result implies that finding steady-state solutions for ODE systems of zero, first and homogeneous second-order chemical reactions is a NP-hard problem; hence, it is not a feasible approach to approximate non-homogeneous second-order reactions into homogeneous ones.
\end{abstract}

\bigskip
 
\begin{tabular}{ll}
{\em Keywords:} & quadratic polynomial system, ordinary differential equation,\\ 
                & chemical reaction kinetics, NP-hard problem
\end{tabular}

\newpage

\section{Introduction}
Let $x_1, \ldots, x_i, \ldots x_n$ be a set of $n$ non-negative, real variables, each one describing the concentration of a chemical species $i$. Consider that two of these chemical species, say $i$ and $j$, $i \ne j$, are in a chemical reaction $k$ such that $i$ is a reactant that is transformed into a product $j$ at a rate constant $\alpha_k$:
\begin{equation} \label{eq:first}
x_i \xrightarrow{\alpha_k} x_j.
\end{equation}
If the rate law of this stoichiometric formula is given by:
\begin{equation} \label{eq:first:rate}
rate^1_k = \alpha_k x_i,
\end{equation}
then say that Equation~\ref{eq:first} describes a first-order reaction. Now, consider that $i$ and $j$ are in a chemical reaction $k$ with two reactants and one product, both reactants being $i$ and the product $j$ being generated at a rate constant $\beta_k$:
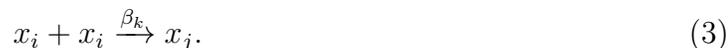
\begin{equation} \label{eq:second}
x_i + x_i \xrightarrow{\beta_k} x_j.
\end{equation}
If the rate law of the chemical reaction is expressed by:
\begin{equation} \label{eq:second:rate}
rate^2_k = \beta_k x_i x_i = \beta_k x_i^2,
\end{equation}
then Equation~\ref{eq:second} describes a homogeneous second-order reaction; the reaction is ``homogeneous" in the sense that the two reactants share the same chemical species. One example of homogeneous second-order reaction is the homo-dimerization, that is, the formation of a dimer from two identical monomers~\cite{moore1999kinetics}.

Now, let us consider a reaction $k$ from either Equation~\ref{eq:first} or~\ref{eq:second}. If the reaction rate of $k$ is independent of the reactant concentration, then we can have a rate law in which the amount of reactant does not matter:
\begin{equation} \label{eq:zero:rate}
rate^0_k = \gamma_k.
\end{equation}
A reaction that follows the rate law described above is called zero-order reaction.

Let us assume a closed chemical system with $m$ reactions, whose reactants and products are composed of the $n$ aforementioned chemical species. For each chemical species $i$, $1 \le i \le n$, let $\mathcal{P}^0_i$, $\mathcal{P}^1_i$, and $\mathcal{P}^2_i$ ($\mathcal{R}^0_i$, $\mathcal{R}^1_i$, and $\mathcal{R}^2_i$) be the sets of zero-order, first-order and homogeneous second-order reactions, respectively, in which $i$ is reactant (product). Thus, we can combine Equations~\ref{eq:first:rate}, \ref{eq:second:rate}, and~\ref{eq:zero:rate} to describe the amount of change in $x_i$ at instant of time $t$, $t \ge 0$, yielding the following ordinary differential equation (ODE):
\begin{equation} \label{eq:ODE}
\frac{dx_i}{dt} = \sum_{l = 0}^{2}{\left( \sum_{k \in \mathcal{P}^l_i}{rate^l_k} - \sum_{k \in \mathcal{R}^l_i}{rate^l_k} \right)}.
\end{equation}
Observe that the right-hand side of Equation~\ref{eq:ODE} is a quadratic polynomial, that is, a polynomial whose degree is at most 2. Moreover, it also holds that all of its monomials are univariate, that is, there is no monomial with two different variables.

Therefore, if we use Equation~\ref{eq:ODE} to define one ODE for each chemical species, we will obtain an ODE system with $n$ equations. Such type of system is intensely studied in many contexts, such as in modeling of cell signaling pathways~\cite{reis2017interdisciplinary}. One property that is often searched on those systems is a steady-state solution, that is, a tuple of initial values for $x_1, \ldots, x_n$ (i.e., values set at $t = 0$) such that there is no change in $x_i$ for all $t > 0$, $1 \le i \le n$. We can look for those steady-state solutions setting each ODE of the system as zero, that is: 
\begin{equation} \label{eq:ODE:steady-state}
\frac{dx_i}{dt} = 0,
\end{equation}
for all $1 \le i \le n$, thus transforming the ODE system into a non-negative, with univariate monomials, quadratic polynomial equation system (NUMQ-PES). Observe that a steady-state solution of the ODE system is equivalent to solve all equations in the related NUMQ-PES problem, that is, to find a root of this polynomial equation system. If there is at least one root, then that system is solvable; otherwise, we say that such system has no solution.

Previous studies focused on the computational complexity of solving general quadratic polynomial equation systems, that is, systems with multivariate monomials and no constraints on their variables. In this type of problem, each polynomial equation $k$, $1 \le k \le m$, has the form:
\begin{equation} \label{eq:general}
\sum_{1 \le i \le j \le n}{\alpha_{i,j}^k x_i x_j} + \sum_{1 \le i \le n}{\beta_{i}^k x_i} + \gamma^k = 0,
\end{equation}
where $x_i$ and $x_j$ are real variables, and $\alpha_{i,j}^k$, $\beta_i^k$ and $\gamma^k$ are real constants. Although a linear polynomial equation system (i.e., a system whose equations have degree at most 1) is solvable in polynomial time (e.g., using an algorithm such as \href{https://en.wikipedia.org/wiki/Gaussian_elimination#Computational_efficiency}{Gaussian elimination}), it is well-known that solving quadratic polynomial equation systems is NP-hard~\cite{dickerson1989functional,patarin1997asymmetric,garey2002computers,tanaka2012efficient}. However, as we discussed above, there is a practical interest in solving the special case of the general quadratic polynomial equation system presented here, the NUMQ-PES problem; therefore, we will in the next section investigate the computational complexity of this special case.

\section{The problem}
In an instance of the NUMQ-PES problem, we have a system of $m$ quadratic polynomial equations, each polynomial equation $k$, $1 \le k \le m$, having the form:
\begin{equation} \label{eq:system}
\sum_{1 \le i \le n}{\left( \alpha_{i}^k x_i^2 + \beta_{i}^k x_i \right)} + \gamma^k = 0,
\end{equation}
where $x_i$ is a real variable, $x_i \ge 0$, and $\alpha_{i}^k$, $\beta_i^k$ and $\gamma^k$ are real constants. Therefore, the central problem studied in this work is the following one.

\begin{prob}{(NUMQ-PES)} \label{prob:opt}
Consider a non-negative, with univariate monomials quadratic polynomial equation system with $m$ equations and $n$ variables. If this system is solvable, then return a root of it; otherwise return NULL (i.e. the system has no solution).
\end{prob}

As we will show in the following, it is unlikely that there is a polynomial-time algorithm for this problem.

\begin{thm} \label{thm:NP-hard}
The NUMQ-PES problem is NP-hard.
\end{thm}

Although it is considered folklore proofs of NP-hardness for special cases of solving quadratic polynomial equation systems, which includes the NUMQ-PES problem, we do not know any explicit reference to a proof of Theorem~\ref{thm:NP-hard}.

Thus, to provide here a self-containing proof of this theorem, we will make use of the satisfiability problem with 3 literals per clause (\href{https://en.wikipedia.org/wiki/Boolean_satisfiability_problem#3-satisfiability}{3-SAT}), a well-known NP-hard (and also NP-complete) problem~\cite{karp1972reducibility}. In the 3-SAT problem, we have a Boolean expression in conjunctive normal form (CNF) with $n$ Boolean variables and $p$ clauses. Each clause has 3 literals, which in turn are composed of either a variable (positive literal) or its complement (negative literal). As an example, we show a 3-SAT instance with 5 variables and 3 clauses:
\begin{equation} \label{eq:example}
(b_1 \vee \neg b_2 \vee b_3) \wedge (b_2 \vee b_3 \vee \neg b_4) \wedge (\neg b_1 \vee b_4 \vee \neg b_5),
\end{equation}
where $\wedge$ is the conjunction operator, $\vee$ is the disjunction operator, and $\neg$ is the negation operator. In this problem, one must decide whether there is a tuple of Boolean values for $b_1, \ldots, b_n$ such that the computation of the whole expression is equal to 1 (i.e., a tuple that satisfies the expression). In the example of Equation~\ref{eq:example}, the assignment $b_1 = 0, b_2 = 1, b_3 = 1, b_4 = 0$, and $b_5 = 0$ satisfies the expression.

\begin{proof}[Proof of Theorem~\ref{thm:NP-hard}]
We will demonstrate that the NUMQ-PES problem is at least as hard as the 3-SAT problem. To this end, we will present a reduction of 3-SAT instances to NUMQ-PES instances; this reduction will be done in a way that a 3-SAT instance is satisfiable iff its equivalent NUMQ-PES instance is solvable. Let us consider a 3-SAT instance with $n$ Boolean variables ($b_1, \ldots, b_n$) and $p$ clauses. We start to construct its corresponding NUMQ-PES instance by defining $n$ equations as the one below:
\begin{equation} \label{eq:binary}
x_i^2 - x_i = 0,
\end{equation}
where $x_i$ is a non-negative real variable that corresponds to $b_i$, $1 \le i \le n$. In the sequence, we will define $p$ equations, one equation per clause. For a clause $j$, $1 \le j \le p$, we design an equation whose left-hand side is defined based on the clause's literals:
\begin{equation} \label{eq:clause}
\sum_{k=1}^3{ l_k^j } - s_j - 1 = 0.
\end{equation}
In the equation above, $s_j \ge 0$ is a slack variable, and $l_k^j$ corresponds to one of the three literals of clause $j$, and is defined as:
\begin{equation} \label{eq:xi}
l_k^j = 
 \begin{cases} 
  x_i,  & \mbox{if the literal $k$ in clause $j$ is $b_i$, $1 \le i \le n$;}\\
  1 - x_i,  & \mbox{otherwise } (\neg b_i).  
  \end{cases}
\end{equation}
To end the presentation of the proposed reduction, let us define how we attribute values for $x_i$, $1 \le i \le n$, from the 3-SAT instance to the reduced NUMQ-PES instance:
\begin{equation}
x_i := 
 \begin{cases} 
  1,  & \mbox{if $b_i = 1$;}\\
  0,  & \mbox{otherwise}.  
  \end{cases}
\end{equation}
Additionally, values for slack variable $s_j$, $1 \le j \le p$, will be assigned in the following way:
\begin{equation} \label{eq:slack}
s_j :=  l_1^j \phantom{.} l_2^j + l_1^j \phantom{.} l_3^j + l_2^j \phantom{.} l_3^j - l_1^j \phantom{.} l_2^j \phantom{.} l_3^j. 
\end{equation}
Observe that, according to Equation~\ref{eq:slack}, $s_j$ is assigned with values in $\{ 0, 1, 2 \}$.

Thus, combining the $n$ equations in Equation~\ref{eq:binary} with the $p$ equations generated with Equation~\ref{eq:clause}, we have $m := n + p$ equations, each one being a polynomial of degree $1$ or $2$ with non-negative real variables and real-valued constants. Therefore, each of these equations can be described through the expression in Equation~\ref{eq:system}, so we conclude that those $m$ equations compose a NUMQ-PES instance. Moreover, this instance was created through a reduction that is linear on $m$, that is, whose complexity is $O(n + p)$, which means it is polynomial on the size of the 3-SAT instance.

Finally, we need to show that a 3-SAT instance is satisfiable iff is respective NUMQ-PES instance, obtained through the aforementioned reduction, is solvable. Let us assume a 3-SAT instance which is satisfiable. Equation~\ref{eq:xi} defines that either $x_i = 0$ or $x_i = 1$, $1 \le i \le n$. Once the roots of Equation~\ref{eq:binary} are precisely $0$ and $1$, all the $n$ equations yielded through Equation~\ref{eq:binary} are solvable whatever are the values of $b_1, \ldots, b_n$. Moreover, consider any $j$ equation, $1 \le j \le p$ among the $p$ equations defined through Equation~\ref{eq:clause}: observe that the first two terms of the left-hand side of $j$ compose a union operation over 3 unit (Boolean) sets:
\begin{equation} \label{eq:union}
l_1^j + l_2^j + l_3^j  - l_1^j \phantom{.} l_2^j - l_1^j \phantom{.} l_3^j - l_2^j \phantom{.} l_3^j + l_1^j \phantom{.} l_2^j \phantom{.} l_3^j.
\end{equation}
Once the clause corresponding to Equation $j$ is satisfiable (initial assumption about the 3-SAT instance), the union of the three literal terms (Equation~\ref{eq:union}) must be equal to $1$. Therefore, we have:
\begin{subequations}
\begin{align}
\sum_{k=1}^3{ l_k^j } - s_j - 1 &= 0\\
l_1^j + l_2^j + l_3^j - (l_1^j \phantom{.} l_2^j + l_1^j \phantom{.} l_3^j + l_2^j \phantom{.} l_3^j - l_1^j \phantom{.} l_2^j \phantom{.} l_3^j) - 1 & = 0\\
l_1^j + l_2^j + l_3^j  - l_1^j \phantom{.} l_2^j - l_1^j \phantom{.} l_3^j - l_2^j \phantom{.} l_3^j + l_1^j \phantom{.} l_2^j \phantom{.} l_3^j - 1 & = 0\\
1 - 1 &= 0\\
0 &= 0.
\end{align}
\end{subequations}
Conversely, let us assume a NUMQ-PES instance as defined by Equations~\ref{eq:binary} and \ref{eq:clause} that is solvable. Once each of the $n$ equations defined through Equation~\ref{eq:binary} has as a solution either $0$ or $1$, the root of this NUMQ-PES instance necessarily is a tuple $x_1, \ldots, x_n, s_1, \ldots, s_p$ such that $x_i \in \{0, 1\}$, $1 \le i \le n$, and $s_j \ge 0$, $1 \le j \le p$. Now we need to show that for each $j$ of the $p$ equations defined through Equation~\ref{eq:clause}, if equation $j$ is solvable then its corresponding 3-SAT clause is satisfiable. Without loss of generality, consider 4 possibilities of values for clauses $l_1^j$, $l_2^j$ and $l_3^j$, and their respective values for $s_j$ to solve $j$: 
\begin{center}
\begin{tabular}{ccc|c|ccc}
\hline
$l_1^j$ & $l_2^j$ & $l_3^j$ & slack variable  & equation $j$ & corresponding clause\\ 
  & & & ($s_j$) value & solvable? & $j$ satisfied?\\ \hline
 $0$ & $0$ & $0$ & None (since $s_j \ge 0$) & No & No\\ 
 $1$ & $0$ & $0$ & $0$ & Yes & Yes\\
 $1$ & $1$ & $0$ & $1$ & Yes & Yes\\
 $1$ & $1$ & $1$ & $2$ & Yes & Yes\\ \hline
\end{tabular}
\end{center}

\bigskip

As it was shown in the table above, a given equation $j$ being solvable implies that its corresponding clause $j$ is also satisfied; once all equations are solvable, the whole 3-SAT instance is satisfiable, thus concluding this proof.
\end{proof}

\section{Conclusion}

In this work, we presented a self-containing proof that to solve a non-negative, with univariate monomials quadratic polynomial system (NUMQ-PES) is a NP-hard problem. This means that it is NP-hard to compute steady-state solutions for ODE systems that describe the dynamics of a set of zero-order, first-order and homogeneous second-order reactions. This fact also implies that, for an ODE system composed of non-homogeneous second-order reactions, to approximate its reactions to homogeneous second-order ones (which could be done if, for each reaction, the concentrations of the two different reactants are similar to each other along the considered time frame) is not a good approach, since the steady-state analysis of the resulting ODE system will ultimately lead to an instance of the NUMQ-PES problem.

\addcontentsline{toc}{section}{Acknowledments}
\section*{Acknowledgments}
We are very thankful to \href{mailto:t.miltzow@gmail.com}{Tillmann Miltzow} for his useful comments and corrections. This work was supported by grant \#13/07467-1, S\~ao Paulo Research Foundation (FAPESP).

\addcontentsline{toc}{section}{References}
\bibliographystyle{unsrt}


\end{document}